\documentclass[aps,pra,onecolumn,superscriptaddress]{revtex4}
\usepackage{graphicx}
\usepackage{gensymb}
\usepackage{amsmath}
\usepackage{amsfonts}
\usepackage{bm}
\usepackage{color}
\usepackage{ulem}
\usepackage{verbatim}
\begin{document}

\title{EPR-based ghost imaging using a single-photon-sensitive camera}


\author{Reuben~S.~Aspden}
\email{r.aspden.1@research.gla.ac.uk}
\affiliation{SUPA, School of Physics and Astronomy, University of Glasgow, Glasgow, G12 8QQ, UK}
\author{Daniel~S.~Tasca}
\email{daniel.tasca@glasgow.ac.uk}
\affiliation{SUPA, School of Physics and Astronomy, University of Glasgow, Glasgow, G12 8QQ, UK}
\author{Robert~W.~Boyd} 
\affiliation{Department of Physics, University of Ottawa, Ottawa, ON K1N 6N5, Canada}
\author{Miles~J.~Padgett}
\affiliation{SUPA, School of Physics and Astronomy, University of Glasgow, Glasgow, G12 8QQ, UK}

--------
\date{\today}


\begin{abstract}
Correlated-photon imaging, popularly known as ghost imaging, is a technique whereby an image is formed from light that has never interacted with the object. In ghost imaging experiments two correlated light fields are produced. One of these fields illuminates the object, and the other field is measured by a spatially resolving detector. In the quantum regime, these correlated light fields are produced by entangled photons created by spontaneous parametric down-conversion. To date, all correlated-photon ghost-imaging experiments have scanned a single-pixel detector through the field of view to obtain the spatial information. However, scanning leads to a poor sampling efficiency, which scales inversely with the number of pixels, N, in the image. In this work we overcome this limitation by using a time-gated camera to record the single-photon events across the full scene. We obtain high-contrast images, $90\%$, in either the image plane or the far-field of the photon pair source, taking advantage of the EPR-like correlations in position and momentum of the photon pairs.  Our images contain a large number of modes, $>500$, creating opportunities in low-light-level imaging and in quantum information processing.
\end{abstract}


\maketitle

\section{Introduction}

In a ghost imaging system the object and the imaging detector are illuminated by two correlated, spatially separated light beams, and the ghost image is obtained from intensity correlations between the two light fields.  These optical correlations can occur in either the quantum or classical regime  \cite{Shapiro:2012}.  Ghost imaging was first demonstrated in 1995 by Pittman {\it et al}.~\cite{Pittman:1995}, who used correlated photon pairs generated by spontaneous parametric down-conversion (SPDC).  They used one of the down-converted fields to illuminate an aperture, and the transmission through it was detected by a non-imaging, single-pixel detector, termed the ÒbucketÓ detector.  An image of the aperture was obtained from the coincidence counts between this bucket detector and a scanning single-pixel detector in the other beam. Bennink {\it et al}.~\cite{Bennink:2002} demonstrated ghost imaging using a classical light source, and Ferri {\it et al}.~\cite{Ferri:2005} have since shown that ghost imaging using classical light sources can match some results thought to be obtainable only when using entangled light.  However, these classical systems cannot produce high-resolution, background-free images derived from both position and momentum correlations within the same source \cite{Bennink:2004}.  Since ghost imaging was first proposed there has been a large body of published work, using both thermal \cite{Bennink:2002, Gatti:2003, Bennink:2004,Gatti:2004PRA,Ferri:2005,Shapiro:2008, Sun:12,Ragy2012}, and quantum entangled light \cite{Bennink:2004,Abouraddy:2001,Erkmen:2008, Osullivan:2010, Jack:2009, Malik:2010} .

\begin{figure}[h]
\begin{center}
\includegraphics[width=13cm]{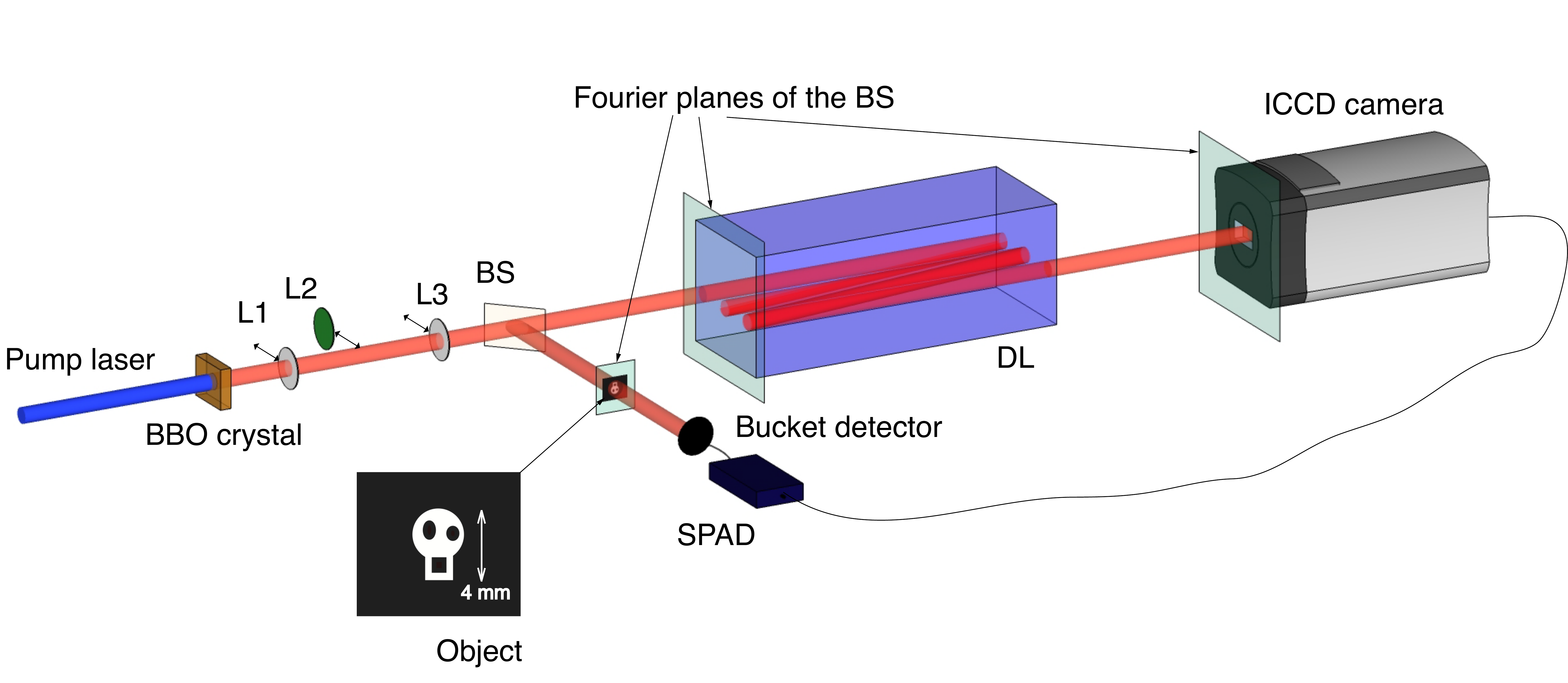}
\caption{Schematic diagram of experimental setup. Collinear down-converted photon pairs at $710$nm are generated by pumping a BBO crystal with a UV laser at $355$nm.  $L_1$ and $L_3$ are $50$mm focal length lenses used to produce an image of the down-conversion source onto the beam splitter (BS).  $L_2$ is a $100$mm focal length lens used to Fourier transform the exit facet of the down-conversion source onto the beam splitter.  A $300$mm focal length lens (not shown) is used after the beam splitter in each path to Fourier transform the down-converted fields at the beam splitter onto the planes of the object in one arm and onto the input plane of an image preserving delay line (DL) in the other arm. The image preserving delay line consists of 7 imaging systems with a total length of $22$m.  Our object (inset) is a transmissive mask with the shape of a skull printed on acetate.}
\label{FIG:setup}
\end{center}
\end{figure}

All previous realisations of ghost imaging using SPDC photons have relied on scanning a detector to measure the spatial distribution of the optical field.  This scanning necessarily limits the detection efficiency of such an imaging system to a maximum of $1/N$, where $N$ is the number of scanned pixels.  In this present work we overcome this limitation and present a practical ghost imaging system at the single-photon level by using a time-gated, intensified CCD (ICCD) camera.  Our advance inherently increases the detection efficiency of a ghost imaging system proportionally to the number of pixels, $N$, in the image.  The use of a camera allows us to capture high-visibility ghost images across the full field of view.  The object is placed in one arm (object-arm) of the SPDC, the transmission through which is detected by a large area bucket detector.  The ICCD is placed in the other arm (camera-arm) and is triggered using the output from the bucket detector, each frame recording the position where the single photon was detected. We compensate for the electronic delay associated with the triggering of the camera by means of a folded, image-preserving delay line as illustrated in figure \ref{FIG:setup}.  

Our system was designed with interchangeable lenses that allow us to utilise either the intensity correlations of the photon pairs in the image-plane of the down-conversion source or their anti-correlation in the far-field.  This capability of using image-plane or far-field correlations allows us to explore the EPR-like correlations in the transverse position or momentum of the photon pairs \cite{Howell:2004}. In our system, the correlations are not those between the camera and the bucket detector, rather, the bucket detector measures the photons whose spatial state is defined by the object, hence the correlations we observe are between the object and the image.

The major advantage of full field-of-view detection, compared to a scanning system, is the dramatic increase in efficiency in the measurement of high-dimensional spatial entanglement. Recent works have demonstrated the potential of the multi-pixel detection of spatial entanglement \cite{Leach:2012, Edgar:2012,Moreau:2012,Dixon:2012,Zhang:2009,Fickler2013} and, albeit for multiple photons, quantum imaging \cite{Brida10}.  In our work, the combination of multi-mode coincidence detection and single-photon sensitivity across an entire field of view suggests applications in low-light imaging systems, and quantum information protocols utilising spatial states such as quantum key distribution \cite{Walborn:2006, Groblacher:2006}, information processing \cite{Tasca:2011} and teleportation \cite{Walborn07}. 

\begin{figure*}[t]
\begin{center}
\includegraphics[width=15cm]{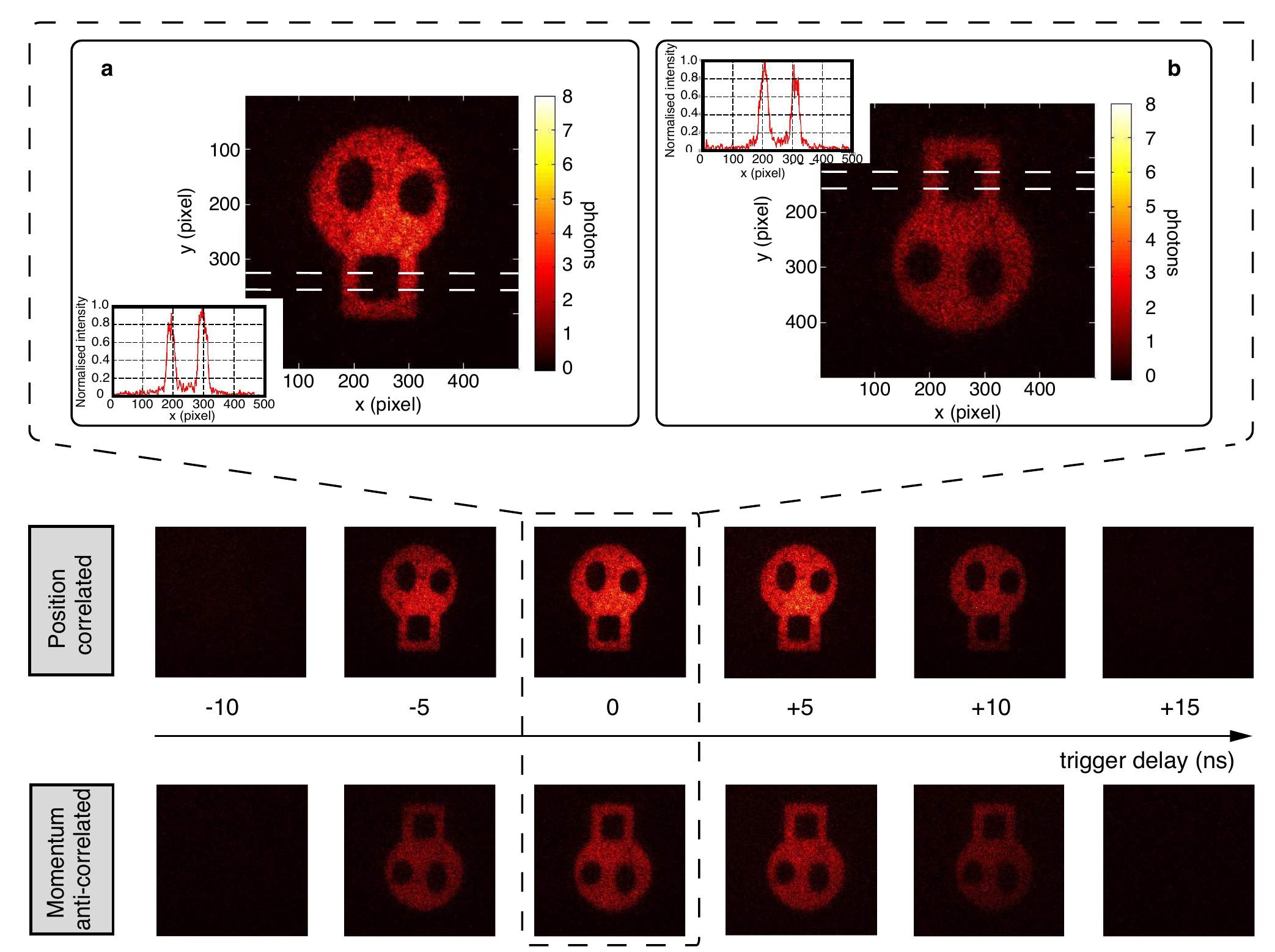}
\caption{Ghost images and trigger delay timing. Ghost images obtained in the position correlated and momentum anti-correlated configurations as a function of the trigger delay, with a total integration time of 1 hour per image. The delay is scanned in increments of $5$ns, which corresponds to adding/removing 1 m of cable. The images shown in (a) and (b) are enlarged versions of the images obtained with optimum delay time, for which the total number of photo events were $124,310$ and $74,021$, respectively. The insets show cross-sections of the images averaged over the 30 rows indicated by the dashed lines.}
\label{FIG:GI}
\end{center}
\end{figure*}

\section{Experimental methods}  A schematic diagram of our experimental setup is shown in figure~\ref{FIG:setup}.  Our down-conversion source consists of a 3-mm-long  $\beta$-barium borate (BBO) crystal cut for type-I phase-matching and pumped with a horizontally polarised, high-repetition rate laser at $355$nm. The pump laser is spatially filtered and re-collimated to a diameter of approximately $1.2$mm (FWHM) at the crystal. The BBO crystal is oriented to provide near-collinear, frequency-degenerate, down-converted photons centred at $710$nm over a $10$nm bandwidth, determined by high-transmission interference filters. We use a pellicle beam splitter (BS) to split the photon pairs into the object and camera arms. As we use a collinear type-I SPDC source, only $50\%$ of the photon pairs are split into different paths at the beam splitter.  The photons used to probe the object are detected with a bucket detector, with a field of view that extends over the whole scene. This bucket detector registers whether the photons have passed through the object whilst recording no direct image information.  Our bucket detector consists of a lens that couples the photons into a multi-mode fibre connected to a single-photon avalanche diode (SPAD).  

Our ghost image is obtained by using the bucket detector to trigger a single-photon acquisition by the ICCD (Andor iStar, Gen 2 image intensifier, WR photo-cathode); we subsequently sum these camera images over many trigger events. The pump power was adjusted to approximately $2$mW, giving a trigger rate from the bucket detector of approximately $15$kHz for the image-plane configuration and $10$kHz for the far-field configuration. This was chosen to be compatible with the maximum response rate of the ICCD whilst minimising the possibility of overlapping trigger pulses. The ICCD was air cooled to $-15^{\circ}$C and operated in direct gate mode, where the TTL pulse from the SPAD triggers the intensifier of the camera directly, giving a gate width equal to the length of the trigger pulse.

To obtain in-focus images, the ICCD camera and the object must be located in planes where the down-converted photons exhibit strong intensity correlations \cite{Tasca:2009B}. We chose two configurations that satisfy this condition.  In the first configuration, both the object and the ICCD are placed in image planes of the BBO crystal, where the positions of the photons are correlated.  In the second configuration, the object and the ICCD are placed in the far-field of the BBO crystal, where the positions of the photons are anti-correlated, arising from their momentum anti-correlation. We choose between these two configurations through use of an interchangeable lens system located between the BBO crystal and the beam splitter, as indicated in figure \ref{FIG:setup}. The ability to show strong correlations in complementary variables, such as in momentum and position, is the hallmark of entanglement and is not possible with a classical source \cite{Howell:2004, Spengler:2012}. 

In order to allow for the electronic delay associated with the bucket detector and the trigger mechanism in the ICCD camera ($\approx 70$ns), the photon in the camera-arm has to be delayed before it arrives at the ICCD.  Our delay line (DL) consists of consecutive imaging systems with unit magnification (see figure \ref{FIG:ExpSetSup}).  There are a total of $7$ such imaging systems in the delay line, $4$ of which are formed with $1000$mm focal length lenses, and $3$ with $500$mm focal length lenses, giving a total distance of propagation of $22$m. Using the aforementioned interchangeable lens system, the two configurations set a relationship between the crystal and object/camera planes that are characterised by either an imaging system with magnification $M = 3$ or a Fourier system with effective focal length $f_e = 300$mm (see \ref{App:Delayline}).


\section{Image acquisition} Figure~\ref{FIG:GI} shows a set of ghost images taken in both the position correlated and momentum anti-correlated configurations for several different values of the delay of the triggering pulse (adjusted by changing the length of the cable between the SPAD and the ICCD).  As the timing of the triggering pulse is changed relative to the time of arrival of the correlated photon on the ICCD, the coincidence is lost and therefore no image is recorded.  Figures~\ref{FIG:GI}-a and~\ref{FIG:GI}-b show the optimum ghost images in the two configurations, both with a contrast in excess of $90\%$. 

All images were obtained by summing $1800$ accumulations of $2$ seconds duration. This accumulation time was chosen in order to ensure that each individual accumulation was spatially sparse, and therefore photon counting could be applied. In this regime it is desirable to have a mean number of photons per mode of much less than one, as the threshold does not distinguish between one and many photons. The resolution of the ICCD is set not by the pixel size but rather by the resolution of the intensifier. Incident photons on the intensifier are converted into photoelectrons, which are multiplied. These photoelectrons fluoresce on a phosphor screen and the photons are detected at the CCD chip. This process sets the resolution and is manifested as a blooming of the photo-events across several pixels on the CCD chip \cite{Buchin:2011}. To compensate for this limited resolution we apply a spatial filter where each event smaller than two pixels in extent is attributed to read-out noise on the chip and is subtracted from the image. Using this photon counting methodology, the average number of detected photons per frame was approximately $70$ in the position correlated configuration and $40$ in the momentum anti-correlated configuration. The intensity scale of the displayed images is the summed number of detected photons per mode.  Importantly, no background subtraction has been applied to any of the images.

In the position correlated configuration the image is upright, figure~\ref{FIG:GI}-a. In the momentum anti-correlated configuration the resulting image is inverted, figure~\ref{FIG:GI}-b. As the optical magnification in each arm is the same, the size of both images on the ICCD match the actual size of the object.  The resolution of both images is comparable and is set by a combination of the correlation length of the down-converted photons and the resolving power of our optical system (see \ref{App:Correlation}).

\begin{figure}[t]
\begin{center}
\includegraphics[width=14cm]{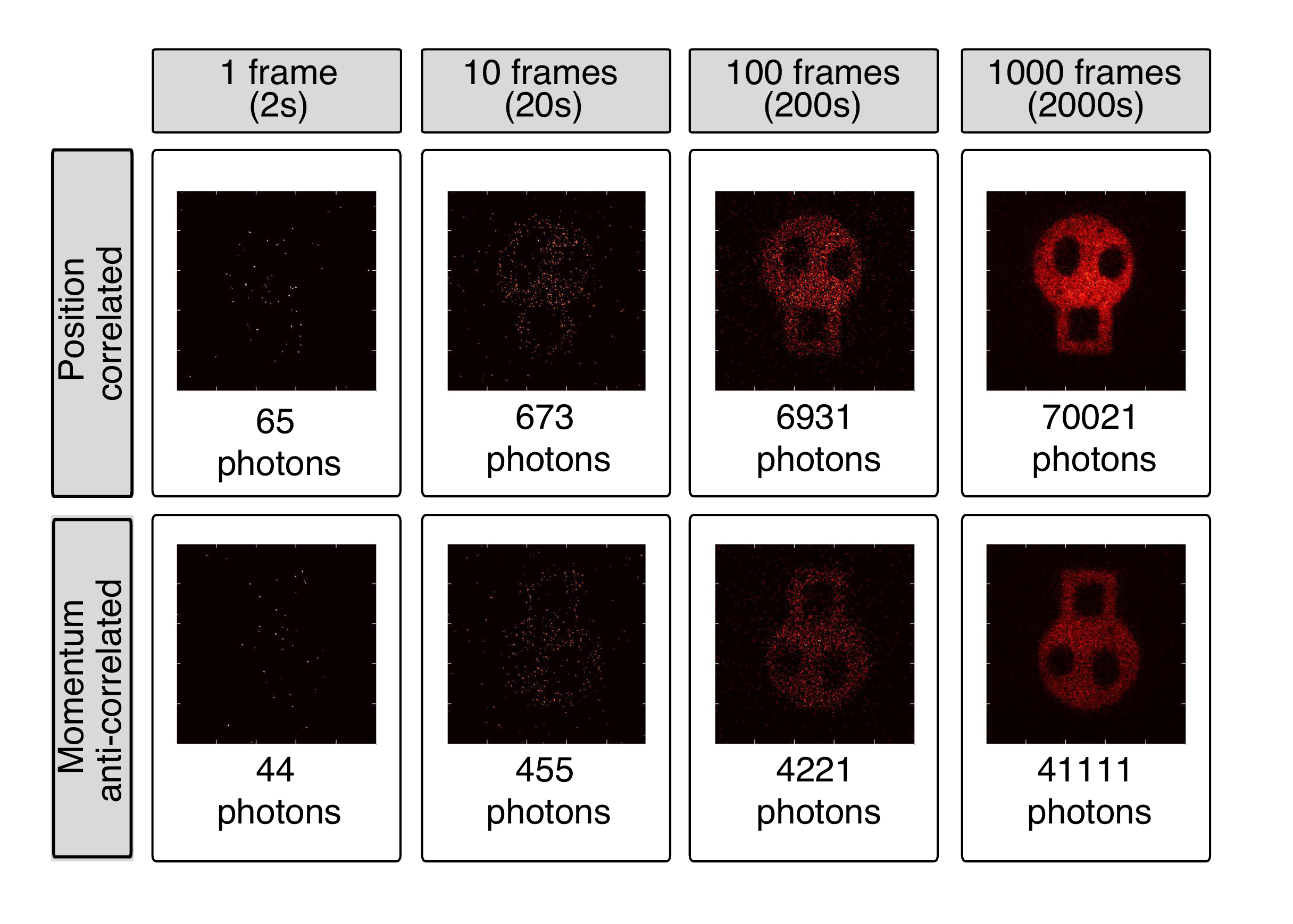}
\caption{ Photon counting and heralding efficiency. Images obtained in position correlated and momentum anti-correlated configurations for an increasing number of acquired frames, along with the corresponding number of detected photons.  Each frame is the result of a thresholding procedure of the recorded intensity distributions over an accumulation time of 2s}
\label{FIG:Photons}
\end{center}
\end{figure}


The ability to obtain high-resolution, background-free images in both the position correlated and momentum anti-correlated configurations is enabled by utilisation of the EPR-like correlations in the spatial variables of the down-converted photons. By analysing the bright-to-dark transitions within the images we obtain point spread functions (PSF) for our ghost imaging system with standard deviations given by $\sigma_{pos} =81\mu$m in the position configuration and $\sigma_{mom}=135\mu$m in the momentum configuration\footnote{Using cross-sections of the images, we calculate the standard deviations of the PSF from the width corresponding to the transition from $90\%$ to $10\%$ of the maximum intensity. This width is equal to $2.56 \sigma$.}. These standard deviations are indicative of the strength of the position and the momentum correlations of the photon fields in the plane of the ICCD and the object. Although not strictly equivalent to a conditional probability distribution across the whole field, they are nevertheless good measures of the strength of the intensity correlations. We estimate the standard deviations of the position, $\Delta_{pos}$, and momentum, $\Delta_{mom}$, correlations at the SPDC source from $\sigma_{pos}$ and $\sigma_{mom}$ and the parameters of the optical systems according to $\Delta_{pos}=\sigma_{pos}/{M}$ and $\Delta_{mom}=(k / {f_e})\sigma_{mom}$. We note that their variance product of $0.012\hbar^2$ is smaller than $\hbar^2/4$.

As an alternative to obtaining ghost images of the object shown in figure \ref{FIG:setup} we also recorded images using pinholes of 75$\mu m$ and $200\mu m$ in diameter. Although of less complexity as an image, the ghost images of the pinholes give the detection probability of the photons in the camera arm conditioned on the measurement of the photons in the object arm within a small spatial extent as defined by the pinhole. These conditional probabilities lead more directly to a measure of the strength of the correlations as  strictly required in a demonstration of EPR correlations {\cite{Reid:1989}. Using these pinholes, we calculated the standard deviations of the PSF in position and momentum to be  $\Delta_{pos}^{x} =(49\pm1)\mu$m, $\Delta_{pos}^{y} =(49\pm3)\mu$m, $\Delta_{mom}^{x}=(128\pm7)\mu$m and $\Delta_{mom}^{y}=(140\pm8)\mu$m giving a variance product of $(3.8\pm0.1)\times10^{-3}\hbar^2$ and $(4.5\pm0.2)\times10^{-3}\hbar^2$ in the x and y directions respectively.  Slight differences between these values and those inferred from the bright-to-dark transitions of the images can be attributed to defocus from the ideal case. However, the strength of the correlations observed in both the printed image and the pinhole are indicative of EPR correlations. 


\section{Information content of the images} The amount of information contained in the transverse distribution of a single photon is an important question in the context of a communication or cryptographic system \cite{Walborn:2006,Walborn:2008,Leach:2012a}.  Two practical limitations to consider when using spatial states in this way are the resolution of the measurement apparatus and the finite size of the transverse field distribution. The resolution of our optical system is set by the PSF, as calculated above. The field distributions in the image-plane and far-field as detected by the camera have a standard deviation of $\Gamma_{pos}=1.83$mm and $\Gamma_{mom}=3.06$mm, respectively. As the object and the ICCD are positioned in equivalent planes, these field distributions correspond to those illuminating the object. The largest dimension of the object is $4$mm (see figure} \ref{FIG:setup}) so it is enclosed within a region smaller than $2.2 \Gamma$ of the illumination beam in both configurations. Considering an information encoding based on binned transverse position states of the single photons, our optical system can distinguish up to $(\Gamma/\sigma)^2$ orthogonal states. Using the PSF and beam widths quoted above, we calculate that our optical system can resolve approximately 500 modes in both configurations. This corresponds to up to $9$ bits of information per detected photon.  Figure~\ref{FIG:Photons} shows a set of images obtained by summing an increasing number of frames. As discussed above, photon-counting was performed by using intensity and spatial thresholding techniques. This gave a heralding efficiency of $\eta=0.2\%$, which was calculated as the ratio between the average number of detected photons at the camera and the triggering rate. This heralding efficiency could be increased by changing the threshold procedures applied, but this would come at the cost of increased dark count rates and reduced visibility. 

\section{Discussion} We have experimentally demonstrated a practical-ghost imaging system at the single-photon level by utilising a multi-pixel, time-gated, intensified camera. We recorded full-field ghost images with $90\%$ visibility by utilising the spatial correlations of photon pairs in position or the anti-correlations in momentum.  The ability to obtain high-contrast images using either position or momentum correlations is an image-based utilisation of the EPR phenomenon.

We circumvent the time delay associated with the trigger mechanism of our camera and other components by building a free-space, image-preserving optical delay line.  The ability to take full-field images gives a potential increase in sampling efficiency which scales with the number of pixels in the image. We believe this ghost imaging system could be useful both in imaging applications as well as in investigations of quantum information protocols with spatial states.  Regarding imaging applications, one could envisage a non-degenerate SPDC system \cite{Rubin:08} where the object is illuminated with infrared light whilst the image is recorded with photons in the visible spectrum.  This non-degenerate approach could be of particular interest for low-light, low-energy imaging of biological samples.

\appendix

\section{Image preserving delay line} \label{App:Delayline} We use a type-I down-conversion source with the optical axis of the BBO crystal orientated horizontally, thus generating vertically polarised down-converted photons. The polarisation of the photons in the camera-arm is rotated by $90^\circ$ using a half-wave plate @ $45^\circ$, so they are transmitted through the polarising beam splitter (PBS). They are then imaged through the PBS to the input mirror of a telescopic imaging system with unit magnification constructed with 1m focal length lenses. The photons propagate through this imaging system and are back-reflected along the same path, double passing through a quarter-wave plate set @ $45^\circ$, thus rotating the polarisation of the photons back to horizontal. They are therefore reflected when incident again on the PBS and are re-imaged on to the camera. 

\begin{figure*}[h]
\begin{center}
\includegraphics[width=16cm]{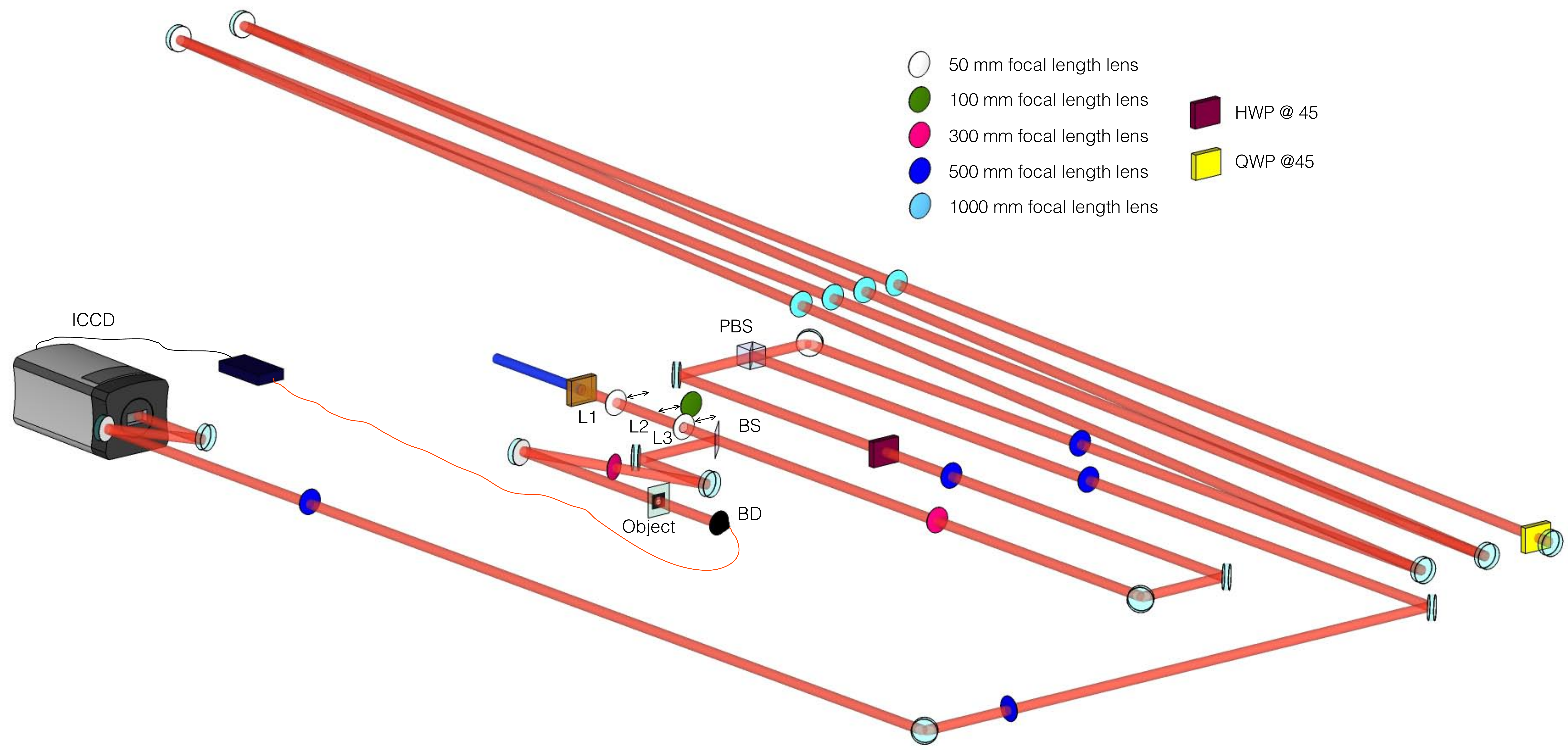}  
\caption{Full experimental setup of our ghost imaging system.}
\label{FIG:ExpSetSup}
\end{center}
\end{figure*}

\section{Two-photon intensity correlation} \label{App:Correlation} In our ghost imaging system, we make use of the intensity correlations of the photon pairs in the image plane and far-field of the down-conversion source. The correlation length of the photon pairs in these planes can be estimated from the two-photon wave function and from our optical system. In the thin-crystal, monochromatic and paraxial approximations, the post-selected two-photon wave function of SPDC can be written as \cite{Monken:1998,Walborn:2010}
\begin{equation}\label{EQ:WFmom}
\Phi(\bold{q}_1,\bold{q}_2)= \mathcal{V}(\bold{q}_1+\bold{q}_2) \gamma(\bold{q}_1-\bold{q}_2),
\end{equation}
where $\bold{q}_1$ ($\bold{q}_2$) is the transverse component of the wave vector of photon 1 (2). The function $\mathcal{V}$ is the angular spectrum of the pump beam, which is transferred to the two-photon wave function \cite{Monken:1998}, and $\gamma$ is the phase-matching function of the SPDC. Since the angular spectrum of the pump beam $\mathcal{V}$ is much narrower than the phase matching function $\gamma$, the photon pairs exhibit anti-correlation between their transverse momenta. In our experiment we use a fundamental Gaussian pump beam such that $\mathcal{V}$ can be approximated by a Gaussian function. The phase matching function, $\gamma$, can be described by a $sinc$ function \cite{Walborn:2010} and under certain conditions is approximated by a Gaussian function \cite{Walborn:2010, Law:2004}.

The two-photon wave function in transverse position representation is given by the Fourier transform of Eq. \eqref{EQ:WFmom}:
\begin{equation}\label{EQ:WFpos}
\Psi(\boldsymbol{\rho}_1,\boldsymbol{\rho}_2)= \mathcal{W}(\boldsymbol{\rho}_1+ \boldsymbol{\rho}_2) \Gamma(\boldsymbol{\rho}_1-\boldsymbol{\rho}_2),
\end{equation}
where $\boldsymbol{\rho}_1$ and $\boldsymbol{\rho}_2$ are the transverse coordinates in the detection planes of photons 1 and 2, $\mathcal{W}$ is the transverse field distribution of the pump beam and $\Gamma$ is the Fourier transform of the phase matching function. As  the function $\Gamma$ is much narrower than $\mathcal{W}$, the transverse positions of the photons are correlated. We define the correlation lengths of the photon pairs as the standard deviation of the two-photon detection probability distributions in image-plane and far-field, which are proportional to the widths of  $|\Gamma|^2$ and $|\mathcal{V}|^2$ respectively.  Using the parameters of our SPDC and optical system, we use the Gaussian approximation \cite{Law:2004} to estimate these correlation lengths as
\begin{eqnarray}
\sigma_{IP} & \approx & \frac{M}{\sqrt{2}}\sqrt{\frac{0.455L}{k_p}}  \approx 19\mu m, \\
\sigma_{FF} &\approx &   \frac{1}{\sqrt{2}}\frac{f_e}{\sigma_{p}k} \approx 33 \mu m,
\end{eqnarray}
where $L=3$mm is the length of the crystal, $k_p$ is the wavenumber of the pump beam, $\sigma_p$ is the standard deviation of the amplitude of the Gaussian pump beam at the crystal, and $f_e$ is the effective focal length of the Fourier system. We note that these correlation lengths are both below the resolution of our optical system given by the widths $\sigma_{pos}=81\mu$m and $\sigma_{mom}=135\mu$m of the PSFs.

{\bf Acknowledgements}

M.J.P. would like to thank the Royal Society, the Wolfson Foundation and DARPA. R.W.B. thanks the DARPA InPho programme.  We acknowledge the financial support from the UK EPSRC. We thank Dr. J. Romero for careful reading of the manuscript.

\end{document}